\documentclass[letterpaper,showpacs,preprintnumbers,amsmath,amssymb,nofootinbib,twocolumn,superscriptaddress,notitlepage,prl]{revtex4-1}
\usepackage{graphicx}
\usepackage{dcolumn}
\usepackage{bm}
\usepackage[usenames,dvipsnames]{color}
\usepackage[utf8]{inputenc}
\usepackage{amsmath}	
\usepackage{amssymb}	
\usepackage{multirow}
\usepackage{gensymb}
\usepackage[colorlinks=true,citecolor=blue,linkcolor=blue]{hyperref}

\def\aap{A\&A}%
\def\pasp{PASP}%

\begin{document}


\title{Imaging Galactic Dark Matter with High-Energy Cosmic Neutrinos}%

\author{Carlos A. Arg\"uelles}
\email{carguelles@fas.harvard.edu}
\affiliation{Department of Physics, Massachusetts Institute of Technology, Cambridge, MA 02139, USA}
\author{Ali Kheirandish}%
 \email{ali.kheirandish@icecube.wisc.edu}
\affiliation{Department of Physics and Wisconsin IceCube Particle Astrophysics Center, University of Wisconsin, Madison, WI 53706, USA}
\author{Aaron C. Vincent}
\email{aaron.vincent@queensu.ca}
\affiliation{
Department of Physics, Imperial College London, London SW7 2AZ, UK
}%

\begin{abstract}
We show that the high-energy cosmic neutrinos seen by the IceCube Neutrino Observatory can be used to probe interactions between neutrinos and the dark sector that cannot be reached by current cosmological methods. The origin of the observed  neutrinos is still unknown, and their arrival directions are compatible with an isotropic distribution. This observation, together with dedicated studies of Galactic plane correlations, suggest a predominantly extragalactic origin. Interactions between this isotropic extragalactic flux and the dense dark matter (DM) bulge of the Milky Way would thus lead to an observable imprint on the distribution, which would be seen by IceCube as 1) slightly suppressed fluxes at energies below a PeV and 2) a deficit of events in the direction of the Galactic center. We perform an extended unbinned likelihood analysis using the four-year high-energy starting event dataset to constrain the strength of DM-neutrino interactions for two model classes. We find that, in spite of low statistics, IceCube can probe regions of the parameter space inaccessible to current cosmological methods.
\end{abstract}

\maketitle

\textit{\textbf{Introduction}} 
Although the effects of cosmological dark matter (DM) have only been observed via its gravitational influence, the order-one ratio between the dark and standard model (SM) baryonic components of the Universe $\Omega_{DM} \sim 5 \Omega_b$ hint at a nongravitational link between the two sectors. An annihilation cross section of DM near the weak scale, for instance, easily produces the observed relic density through thermal decoupling -- an observation known as the \textit{WIMP miracle}\footnote{Though a weak scale is not even necessary, see e.g. the \textit{WIMPless miracle}~\cite{Feng:2011in}.} (see e.g.,~\cite{Jungman:1995df}). Regardless of the exact production mechanism, the existence of an annihilation process ${\rm DM} \rightarrow {\rm SM}$ implies that the DM-SM elastic scattering process exists and may be measured in the laboratory. Most notably, this is the basis for the plethora of underground direct detection experiments, that aim at observing elastic scattering between DM and quarks. However, interactions between the DM and other particles may also be present and could be the dominant, or even sole, link between the dark and visible sectors. A DM-neutrino interaction is especially attractive for light DM models, where annihilation into heavier products is kinematically forbidden, and appears naturally in some models, for example when the DM is the sterile neutrino (see \cite{Adhikari:2016bei}).  

Such a possibility has been considered in depth, mainly in the context of cosmology~\cite{Boehm:2000gq,Boehm:2001hm,Boehm:2004th,Bertschinger:2006nq,Mangano:2006mp,Serra:2009uu,
Wilkinson:2014ksa,Aarssen:2012fx,Farzan:2014gza,Boehm:2014vja,Cherry:2014xra,Bertoni:2014mva,Schewtschenko:2014fca}. Two main effects have been considered. First, light ($\lesssim 10$ MeV) DM can transfer entropy into the neutrino sector as it becomes nonrelativistic, affecting the expansion rate of the Universe and dramatically altering Big Bang nucleosynthesis  (BBN) and cosmic microwave background (CMB) observables. Second, a small ongoing interaction will lead to \textit{diffusion damping} of cosmological perturbations on small scales, as power is carried away from the collapsing DM overdensities. This suppresses the matter power spectrum and CMB structure on small scales, allowing limits on the nonrelativistic DM-neutrino cross section to be placed.

In this work we turn to a novel, complementary approach. We focus on present-day interactions between high-energy cosmic neutrinos and the DM halo of the Milky Way.  After four years of data taking, the IceCube Neutrino Observatory has confirmed the observation of 53 high-energy contained-vertex events (or ``High-Energy Starting Events, HESE"Â). Analyses of these events  reject a purely atmospheric origin at more than 6$\sigma$ and show compatibility with an isotropic distribution~\cite{Aartsen:2015zva}. This isotropy has been used to place constraints on a Galactic contribution, either from standard sources~\cite{Ahlers:2015moa} or from the decay~\cite{Bai:2013nga} or annihilation~\cite{Zavala:2014dla} of halo DM. Here, we search for DM-neutrino interactions by their effect on the isotropy of the extragalactic signal. As they pass through the Milky Way on their way to Earth, the flux of neutrinos would be preferentially attenuated in the direction of the Galactic center, where the DM column density is the largest. For large enough coupling strengths, this leads to an observable, energy-dependent anisotropy in the neutrino sky.\footnote{We note that complementary information can be gained from high-energy neutrino scattering with the cosmic relic neutrino background \cite{Cherry:2014xra,Cherry:2016jol}.}  

We begin by describing the DM-neutrino interactions studied in this work; this is followed by a description of the IceCube events and of the physical model and likelihood that we consider. Finally, we present constraints on the parameter space of such models using IceCube four-year HESE dataset. \\

\textit{\textbf{Neutrino-dark matter interaction}}
At low energies, when the neutrino energy is much less than the DM mass, elastic scattering cross sections between a neutrino $\nu$ and a heavy particle $\chi$ typically scale as $\sigma \propto E_\nu^2$. Large-scale structure surveys provide the strongest constraint on such interactions due to diffusion damping of primordial fluctuations, as they ``bleed'' power into the relativistic degrees of freedom: $\sigma < 10^{-40} (m_\chi/\mathrm{GeV})(T/T_{0})^2$ cm$^2$, where the neutrino temperature today is $T_0$ (in the case of constant cross section, $\sigma < 10^{-31}(m_\chi/\mathrm{GeV})$ cm$^2$)~\cite{Escudero:2015yka}. Concurrently, thermal contact between neutrinos and DM after neutrino decoupling ($T_{dec} \simeq 2.3$ MeV) leads to injection of entropy into the neutrino sector as the DM becomes nonrelativistic, which affects both nucleosythesis and recombination by accelerating the expansion rate of the Universe. Based on recent CMB measurements and primordial elemental abundance determinations, this second, independent effect leads to the constraint $m_\chi \gtrsim 4$ MeV for a real scalar and $m_\chi \gtrsim 9$ MeV for Dirac fermionic DM~\cite{Wilkinson:2016gsy}.

At high energies, the approximation $\sigma \propto E_\nu^2$  breaks down for most viable particle interactions, since any mediating particle $\phi$ starts to be resolved as the center of mass energy approaches $m_\phi$. We thus turn to two simplified interaction models to illustrate our scenario: a) a fermionic DM candidate coupled to neutrinos via a spin-1 mediator $(S_\chi,S_\phi) = (1/2,1)$ and b) a scalar DM particle, coupled via a fermion $(0,1/2)$. The former is akin to a new $Z'$ gauge boson~\cite{Boehm:2002tm}, while the latter, inspired by right-handed sneutrino models (e.g.,~\cite{Cerdeno:2008ep,Cerdeno:2009dv}), includes an s-channel diagram in the elastic scattering matrix element and thus presents some resonant structure. This leads to qualitatively different phenomenology, suggesting that resonant scattering at high energies may be significantly more constraining than cosmological constraints where $E_\nu \lesssim$ eV $\ll m_\chi$. 

In both cases, we refer to the DM as $\chi$ and the mediator as $\phi$, and seek to constrain the particle masses $m_\chi, m_\phi$, and three-point couplings $g$ (setting the $\phi-\nu$ and $\phi-\chi$ couplings equal where relevant). \\

\textit{\textbf{The extragalactic neutrino signal}}
Since the discovery of cosmic neutrinos in 2013~\cite{Aartsen:2013jdh, Aartsen:2014gkd}, IceCube has reported 53 HESE events. Several scenarios and source classes have been proposed for the origin and production of high-energy neutrinos (see~\cite{Padovani:2015mba, Krauss:2014tna, Tjus:2014dna,Hooper:2016jls,Stecker:2013fxa,Loeb:2006tw,Murase:2013rfa,Bai:2013nga}). However, the sources of IceCube's highest energy neutrinos are still a mystery, since -- so far -- all point source searches and correlation studies have favored an isotropic distribution~\cite{Aartsen:2014cva}. This, along with the relatively large observed flux, implies that a large fraction of the energy in the nonthermal Universe originates from hadronic processes. The observed cosmic neutrino flux is predominantly extragalactic in origin, and its total energy density is similar to that of photons measured by the Fermi gamma ray telescope~\cite{halzen:2016}. This suggests a common origin of high-energy neutrinos and gamma rays. That is, rather than some exotic sources, IceCube is observing the same Universe astronomers do.

 In this work, we use the full four-year HESE sample, which consists of 13 muon tracks and 40 cascades. They are compatible with a power law spectrum given by $E^2 \Phi(E) = 2.2\pm0.7 \times 10^{-8}(E/100\,\mathrm{TeV})^{-0.58}\mathrm{\,GeV \, cm^{-2}\, s^{-1}\, sr^{-1}}$.
No statistically significant clustering has been found in this event selection, i.e., the spatial distribution is consistent with being isotropic. Furthermore, correlation between the neutrinos arrival directions and the galactic plane was not found to be significant~\cite{Aartsen:2015zva}. 
\begin{figure}
\centering\leavevmode
\includegraphics[width=\columnwidth]{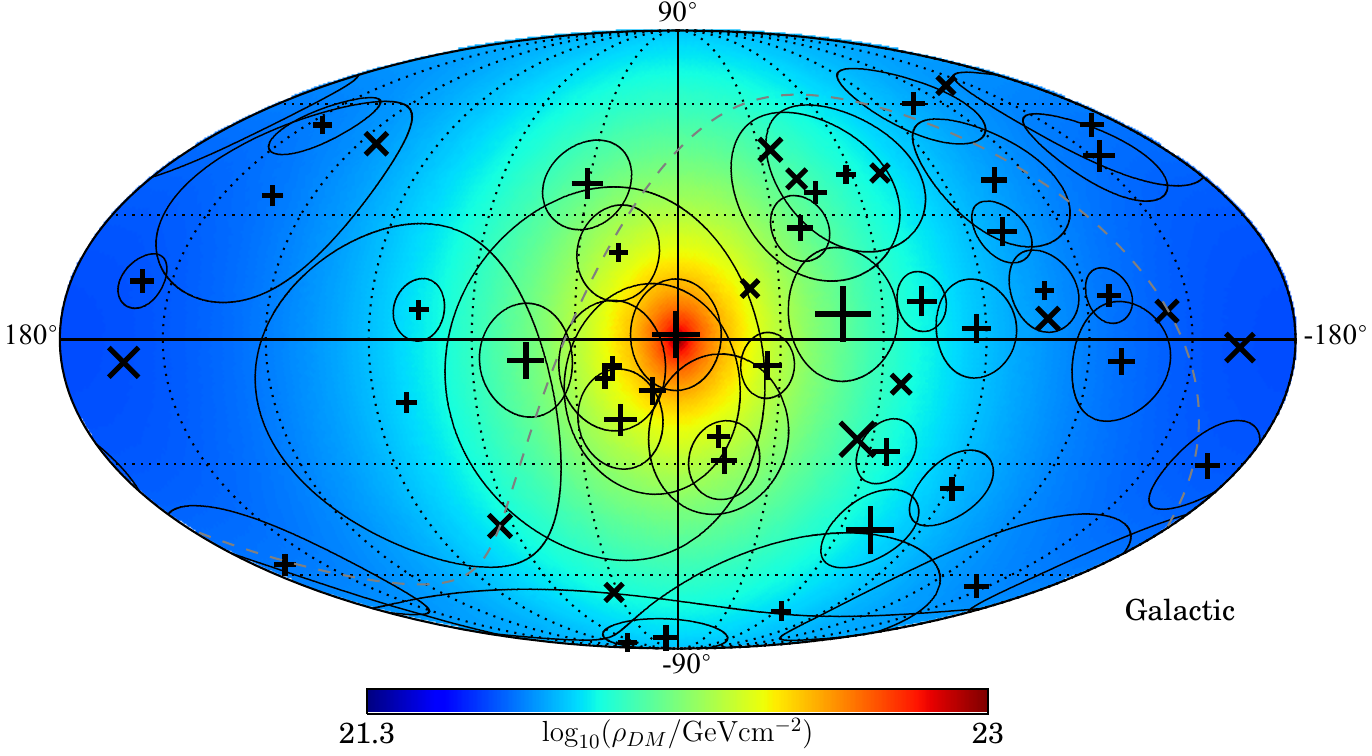}
\caption{The arrival directions of the 53 HESE neutrinos observed in four years of IceCube data~\cite{Aartsen:2015zva}, in Galactic coordinates. Crosses represent shower events, while x's correspond to tracks. Symbol size is proportional to the event energy, and the circles represent the median angular uncertainty of cascades. The color scale is the column density of DM traversed by neutrinos arriving from each direction.}
\label{fig:skymap}
\end{figure}
The flavor composition of this sample is consistent with $(\nu_e:\nu_\mu:\nu_\tau) = (1:1:1)$~\cite{Aartsen:2015knd,Vincent:2016nut}. This is the composition expected for pionic origin of the events and current measured neutrino mixing angle~\cite{Gonzalez-Garcia:2015qrr}. Nonetheless, a different flavor composition at production would yield an oscillation-averaged flux that is very close to $(1:1:1)$ and, with current statistics, would not be distinguishable within the space of flavors allowed by oscillation~\cite{Vincent:2016nut}. In fact, as long as the production mechanism is pion-dominated, the expected flavor ratio remains close to $(1:1:1)$ even in the presence of new physics in the propagation~\cite{Arguelles:2015dca}. We will consider spectral indices of the astrophysical flux between $\gamma = 2$ (corresponding to the expected value from Fermi acceleration) through to $\gamma = 2.9$, consistent with the best fit to the latest HESE data \cite{Vincent:2016nut,ClaudioTalk}. 

We model the attenuation of extragalactic neutrinos as they pass through the halo of DM particles that gravitationally bind the Milky Way. The bulk of the DM lies in the direction of the Galactic center, $(l,b) = (0,0)$ in Galactic coordinates, 8.5 kpc away from our location. Its radial density distribution $\rho_\chi(r)$ can be modeled with the Einasto profile~\cite{1989A&A...223...89E}. We employ  shape parameters that fit the Via Lactea II simulation results~\cite{2009Sci...325..970K} ($\alpha = 0.17, R_s = 26$ kpc), and a local DM density of $\rho_\odot = 0.4$ GeV cm$^{-3}$. A ``cored'' profile $(\alpha = 0.4)$ only leads to slightly less suppression in the very center for a range of cross sections; these to not significantly impact the observables, as they would e.g. for DM annihilation, which depends on the square of the DM density profile.

We take the incoming differential neutrino flux, $\Phi(E)$, to be isotropic. This is not an assumption that all sources are the same: it is rather the statement that in any given direction, the sum of contributions from neutrino sources along the line of sight is the same as from any other direction. We model $\Phi(\tau = 0)$ as a power law in energy. The propagation of the extragalactic high-energy neutrino flux towards the Earth, as they traverse the diffuse DM halo, can be described by a cascade equation
\begin{equation}
\frac{d\Phi(E)}{d\tau} = -\sigma(E)\Phi(E) + \int_E^\infty d\tilde E \frac{d\sigma(\tilde E,E)}{d E}\Phi(\tilde E), \label{eq:cascade}
\end{equation}
where $E$ is the neutrino energy. $\sigma(E)$ is the model-dependent total cross section of $\nu$ with energy $E$, while $d\sigma(\tilde E,E)/dE$ is the differential cross section from $\tilde E$ to E. $\tau$ is the DM column density
\begin{equation}
\tau(b,l) = \int_{\rm{l.o.s.}} n_\chi(x;b,l)~ dx,
\label{eq:coldens}
\end{equation}
$b$ and $l$ are respectively the galactic latitude and longitude, and $n_\chi(x; b,l) = \rho_\chi(r)/m_\chi$ is the DM number density along the line of sight (l.o.s). The DM column density and the arrival direction of high-energy cosmic neutrinos are shown in Fig \ref{fig:skymap}.\\
 
\begin{figure}
 \includegraphics[width=0.5\textwidth]{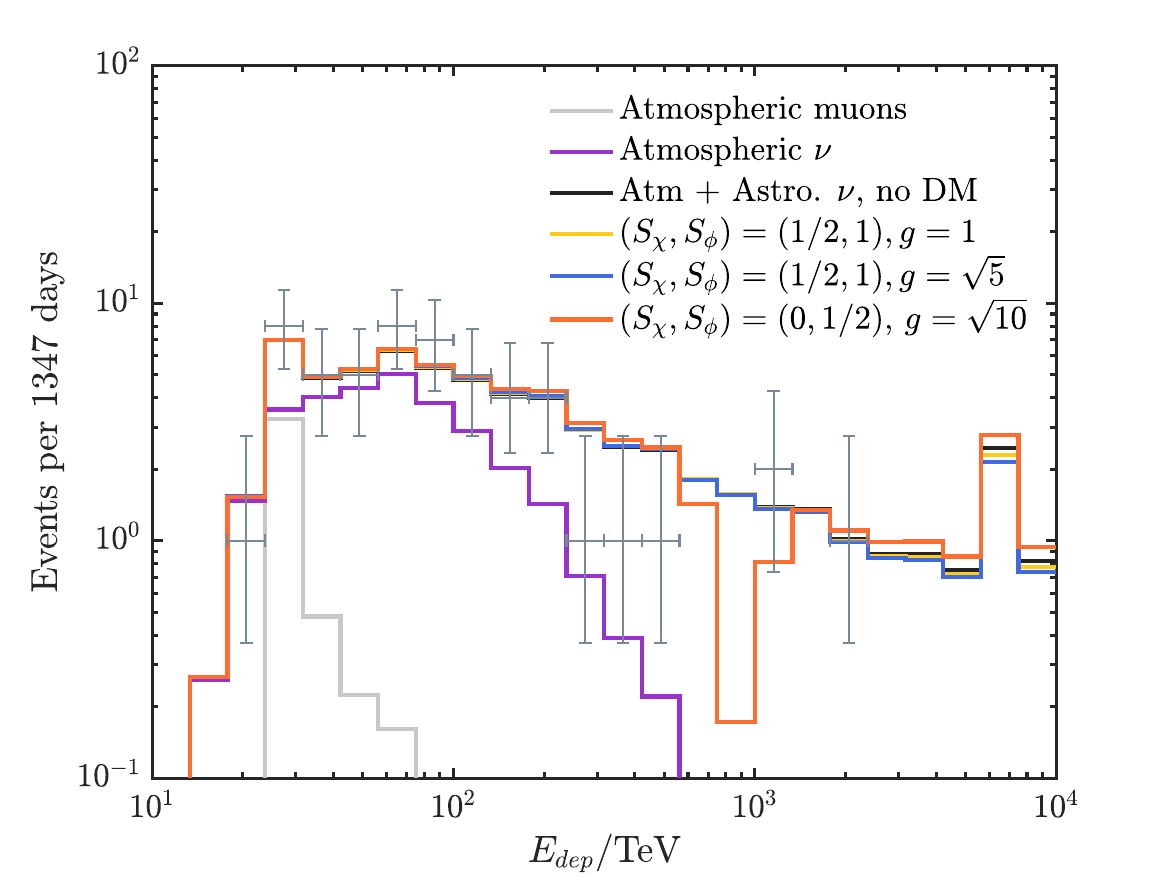}\\
 \includegraphics[width=0.5\textwidth]{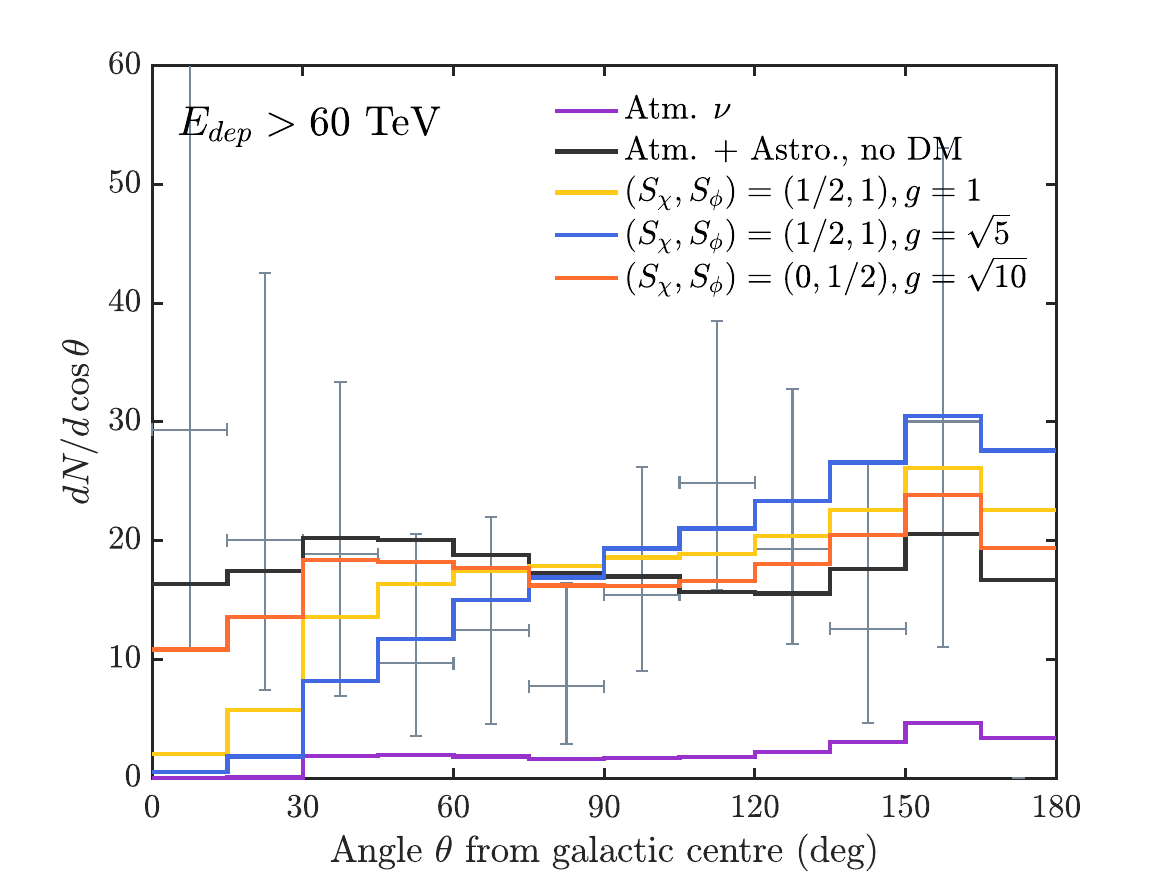}\\
 \caption{Effect on the energy and spatial distribution of HESE as seen at IceCube, due to interactions with the DM halo of the Milky Way for three different examples representative of the parameter space explored in this study. Pale grey and purple lines represent atmospheric background fluxes. Darker lines are: Black: standard astrophysical flux; yellow: fermionic DM with a spin-1 mediator ($g = 1, m_\chi = 10$ MeV, $m_\phi = 10$ MeV). Blue: the same model but with $g = \sqrt{5}$, $m_\chi = 100$ MeV; and orange: scalar DM with a fermionic mediator ($g = \sqrt{10}, m_\chi = 20$ keV, $m_\phi = 6$ GeV). The new physics models can be probed with our analysis of HESE neutrinos, but are not accessible to cosmological studies. We show binned IceCube HESE data as gray crosses.}
 \label{fig:validation}
 \end{figure}
\textit{\textbf{Likelihood function}} We construct an extended unbinned likelihood function for a given set of parameters $\vartheta=\{m_\chi,m_\phi,g \}$ and events of observed topologies $t$, energies $E$, and arrival directions, $\vec x = (b,l)$ 
\begin{equation}
\hspace{-0.30cm} \mathcal{L}(\{t,E, \vec x\}| \vartheta) = e^{-\sum_{b} N_{b}} \prod_{i=1}^{N_{\rm{obs}}}\sum_{a} 
 N_{a} P_{a}(t_i,E_i,\vec x_i| \vartheta),
 \label{eq:likedef}
\end{equation}
where the indices $a, b$ run over the number of astrophysical events ($N_{\rm{astro}}$), atmospheric neutrinos ($N_{\rm{atm}}$), and atmospheric muons ($N_{\mu}$) in the model; while the product in $i$ runs over the observed events ($N_{\rm{obs}} = 53$). The probability of the astrophysical component is proportional to the solution $\Phi(E,b,l)$ of Eq. \eqref{eq:cascade}. A suppression from dark matter in the extragalactic neutrino flux from the $(b,l) = (0,0)$ direction thus suppresses the likelihood of observing astrophysical events from that direction. The probability distributions of the neutrino components in Eq. \eqref{eq:likedef} are given in Appendix A of Supplemental material \cite{supp}. \\

\textit{\textbf{Results}} The likelihood is incorporated into a custom-built Markov Chain Monte Carlo (MCMC) code\footnote{We use the publicly available \texttt{emcee}~\cite{2013PASP..125..306F} sampler.}, which is used to produce posterior likelihood distributions in the six-dimensional space of ($g, m_\chi, m_\phi, N_{\rm{astro}},N_{\rm{atm}},N_\mu$). We note that posteriors on $\{N_a\}$ reproduce independently obtained results~\cite{Aartsen:2015zva,Vincent:2016nut}, with $N_{\rm{astro}} = 34.3\pm 6.5, N_{\rm{atm}} = 14.4 \pm 4.6$, and $N_\mu = 7.1\pm 2.8$. We find that these are completely uncorrelated with the other model parameters. 

Fig. \ref{fig:validation} shows examples of the event distributions in four different scenarios, as they would be expected in IceCube, in the case of an $E^{-2}$ diffuse isotropic flux. The top panel shows the deposited energy distribution, while the lower one shows the event rate versus angular distance from the Galactic center, where DM-induced attenuation is strongest.  Fig. \ref{fig:validation}, highlights the two main effects we observe: 1) a suppression of the event rate as a function of energy, and 2) a suppression of the event rate near the Galactic center. It is the combination of these effects that constrain such models. In the bottom panel of Fig. \ref{fig:validation} we show only events with energies above 60 TeV to avoid confusion with the atmopheric contamination, which comes predominantly from low energies and low declinations.

We contrast four different examples: 1) a null isotropic (black) hypothesis where no DM-neutrino interaction is present, 2) the case (yellow) with a fermionic ($S_\chi = 1/2$, $m_\chi = 10$ MeV) DM particle, with a vector ($S_\phi = 1$, $m_\phi = 10$ MeV) mediator and a coupling g = 1; 3) the same (blue), but with  larger $g = \sqrt{5}$ and mass $m_\chi = 100$ MeV, and finally 4) a scalar DM candidate with a fermionic mediator ($S_\chi = 0, S_\phi = 1/2$), and with $m_\chi = 20$ keV, $m_\phi =  6$ GeV, and $g = \sqrt{10}$.  For reference, we also show atmospheric muons (grey) and neutrinos (purple).
 
Models 2 and 3 are chosen to give an observable effect: these are large, and we will show that they are excluded by our analysis. However, the resulting low-energy cross sections are not large enough to affect the cosmological limits. Indeed,  even the more extreme (blue) of these scenarios remains two orders of magnitude below the large-scale structure limits, and it is therefore clear that in certain regions of the parameter space, IceCube data can provide strong constraints on new $Z'$-like mediators. The fourth and final scenario is chosen such that resonant (s-channel) scattering occurs at $E_\nu =  m_\phi^2/(2m_\chi) = 810$ TeV, close to where the event ``gap'' between \~400 TeV to ~1 PeV, in the observed IceCube events' energy distribution (see crosses in the upper panel of Fig. \ref{fig:validation}). This resonant suppression can clearly be seen in the orange line. However, we note that although the cross section is allowed by diffusion damping constraints, the DM masses required for such an effect are so low that their thermal contact with neutrinos at early times will inevitably increase the number of effective relativistic degrees of freedom, affecting nucleosynthesis and recombination.
\begin{figure}[t]
\centering\leavevmode
\includegraphics[width=0.5\textwidth]{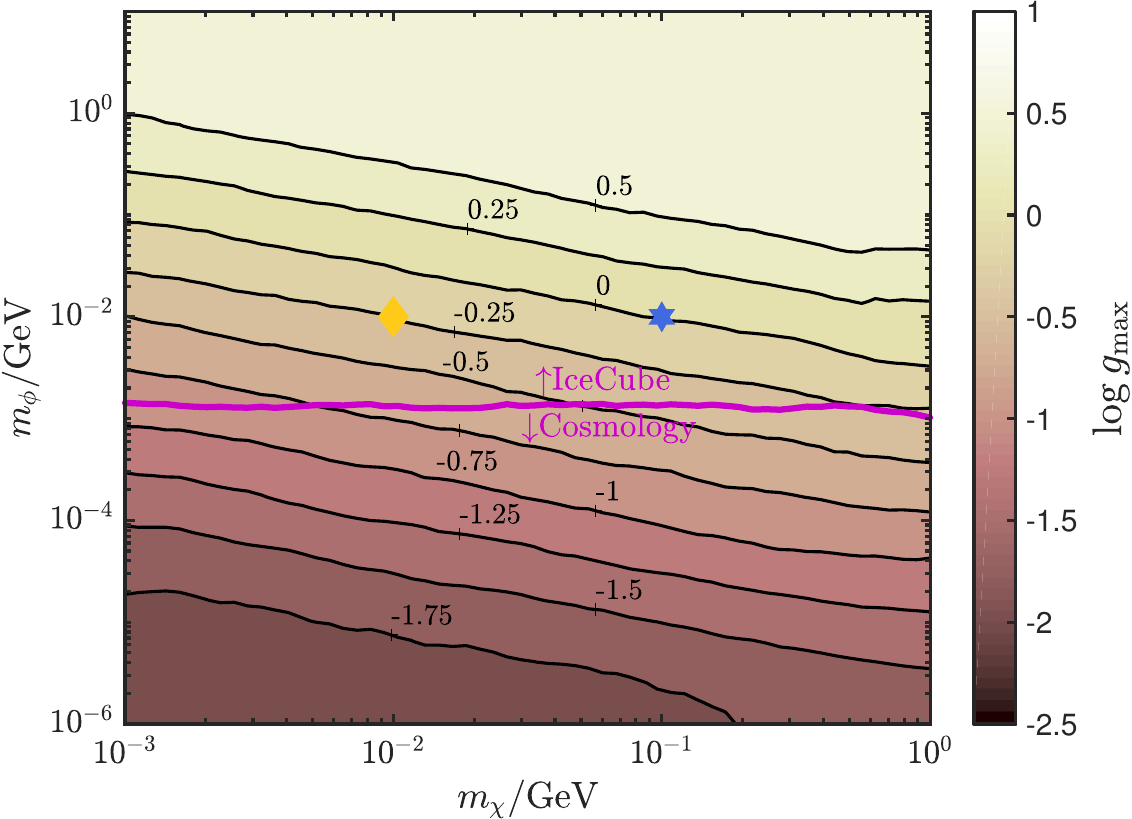}\\
\includegraphics[width=0.5\textwidth]{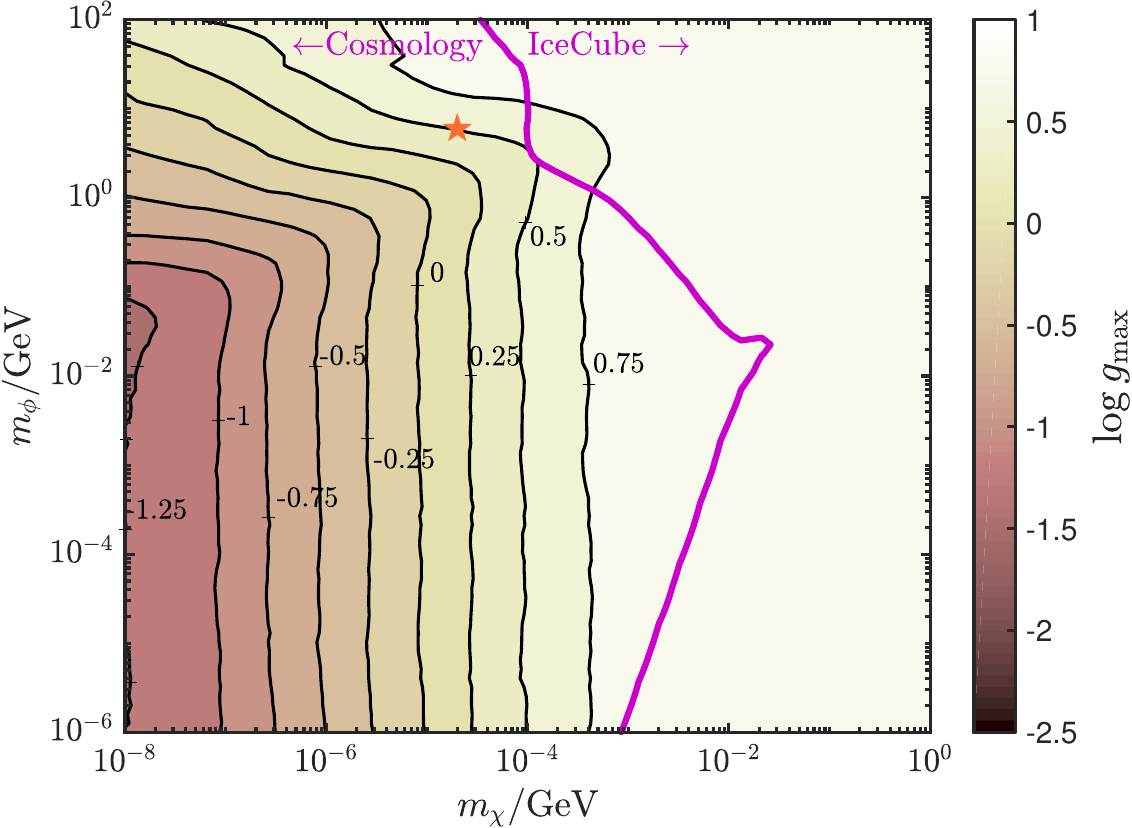}
\caption{Contours of the maximum value of the coupling (shown as $\log g$) allowed by IceCube data, as a function of the dark matter mass $m_\chi$ and mediator mass $m_\phi$. Top: fermionic dark matter coupled to neutrinos via a vector mediator.   Bottom: Scalar dark matter, coupled via a fermionic mediator. In each panel, the thick purple line indicates whether cosmological or IceCube limits are strongest. Examples 2 and 3 from Figure \ref{fig:validation} are shown as a yellow diamond and blue star, respectively. Their coupling lies above the maximum value for their location on the plot. Example 4 is shown in the bottom panel, as an orange star. In this case, cosmology is providing the strongest limits.
}
\label{fig:results_nores}
\end{figure}

In Fig. \ref{fig:results_nores}, we show the results of our MCMC exploration of the full parameter space, marginalized over possible astrophysical and atmospheric fluxes, including values of the astrophysical spectral index between $\gamma = 2$ and $\gamma = 2.9$.\footnote{Appendix B of Supplemental Material \cite{supp} shows the small effect of fixing the spectral index.} Each panel shows contours of the maximum allowed value of the coupling $g$, as a function of the dark matter and mediator masses. The top panel corresponds to the vector mediator case. Above the purple line, our IceCube exclusions are more sensitive than bounds from large scale structure. The bottom panel shows the same parameter space, for the scalar DM model. Here, IceCube fares better on the right-hand side of the purple line. Of particular interest are the distinct resonance regions: from cosmology, this occurs at $m_\phi \rightarrow m_\chi$; in the case of galactic dark matter, the enhancement (and hence, stronger constraints) occurs for $m_\phi^2/2m_\chi \sim 0.1-1$ PeV. We note that variations of the astrophysical spectral index affect constraints near the edges of the region under investigation. However, it is important to note that they do not affect the location of the line beyond which IceCube constraints become stronger than those from large scale structure.\\

\textit{\textbf{Conclusions}} 
We have used the recently opened cosmic neutrino frontier to seek out new dark sector interactions, and to boldly explore the parameter space of their interactions with high-energy neutrinos.  The isotropic distribution of the high-energy extraterrestrial neutrinos leads to constraints on the DM-neutrino interaction in the Galaxy. Our study considers not only the spatial component, but also the energy and topology of the events. We obtain the strongest constraints to date in some regions of parameter space, and our new method is sensitive to the DM-neutrino interaction details. Meson decay experiments have constrained similar interactions \cite{Pasquini:2015fjv, Bakhti:2017jhm}, but not, as yet, for the models considered here, nor in a flavor-blind way. 

Our current constraints are statistically limited due to the low rate of cosmic neutrino detection and the poorness of angular resolutions. Further observation of high-energy starting events in IceCube and implementation of a new method for increasing the number of astrophysical neutrinos from the southern sky would help to obtain stronger bounds or find evidence of DM-neutrino interaction. This also motivates an IceCube extension such as purposed IceCube-Gen2 \cite{Aartsen:2014njl} as well as KM3NeT~\cite{Adrian-Martinez:2016fdl} which will provide larger samples of well reconstructed tracks, and provide more insight into the nature of DM-neutrino interactions. \\ 

\textbf{\textit{Acknowledgements}} We thank Janet Conrad and Francis Halzen for useful comments and discussions. We also thank Markus Ahlers, Ben Jones, Rachel Carr, Alex Geringer-Sameth, Jordi Salvado, Pat Scott, and Donglian Xu. CA is supported by NSF grants No. PHY-1505858 and PHY-1505855.  AK was supported in part by NSF under grants ANT-0937462 and PHY-1306958 and by the University of Wisconsin Research Committee with funds granted by the Wisconsin Alumni Research Foundation. ACV is supported by an Imperial College Junior Research Fellowship.

\bibliographystyle{apsrev4-1}
\bibliography{DarkIce}

\appendix
\onecolumngrid
\newpage
\section{Appendix A: Construction of likelihood function}
Here, we provide the individual contributions to the full likelihood function defined in Eq. (3) 
The likelihood that a given event $i$ is astrophysical is the following
\begin{eqnarray}
P_{{\rm{astro}},i} = \frac{\sum_{f = e,\mu,\tau} P_{\rm{t}}(f|t_i)\int dE  A_{\rm{eff}}^f(E,\vec{x}) R(E,E_{\rm{dep}}E_{\rm{dep}})  \Phi_{\rm{astro}}}{\int dE_{\rm{dep}} \int d^2\vec{x}\sum_{f = e,\mu,\tau} P_{\rm{t}}(f|t_i)\int dE  A_{\rm{eff}}^f(E,\vec{x}) R(E,E_{\rm{dep}})  \Phi_{\rm{astro}}},
\label{eq:pastro}
\end{eqnarray}
where $P_{\rm{t}}$ is the probability that the observed topology $t =$ (track, shower) originated from incoming neutrino flavor $f$. We introduce the detector energy resolution $R_E$ as normal distributions with uncertainties given in~\cite{Aartsen:2015rwa}. The neutrino effective areas for each flavor, $A_{\rm{eff}}^f(E,\vec{x})$, parametrize the detector geometry as well as the neutrino-nucleus (and neutrino-electron) cross sections and the effects of attenuation from propagation within the earth, atmosphere and ice sheet; they thus depend on energy and zenith angle~\cite{IceCubeScience}. $\Phi_{\rm{astro}}$ is the solution to the cascade equation (1) and represents the incoming neutrino flux at Earth, after DM-neutrino interactions have taken place. It depends on neutrino energy and arrival direction $\vec x = (b,l)$. The trivial solution (where $\sigma_{\nu-DM} =0$, or $\tau = 0$) is an isotropic power law in energy. The denominator of Eq. \eqref{eq:pastro} ensures the likelihood is normalized to one. The integral over $E_{\rm{dep}}$ goes from 10 TeV to 10 PeV; this means that the non-observation of events in energy ranges where a $P_{\rm{astro}}$ is high is implicitly taken into account in the likelihood.  

The atmospheric likelihood per event is the following
\begin{eqnarray}
P_{{\rm{atm}},i} = \frac{\sum_{f = e,\mu} P_{\rm{veto}} P_{\rm{t}}(f|t_i)\int dE  A_{\rm{eff}}^f(E,\vec{x}) R_E(E,E_{\rm{dep}})  \Phi_{\rm{atm}}}{\int dE_{\rm{dep}} \int d^2\vec{x}\sum_{f = e,\mu,\tau} P_{\rm{veto}} P_{\rm{t}}(f|t_i)\int dE  A_{\rm{eff}}^f(E,\vec{x}) R(E,E_{\rm{dep}})  \Phi_{\rm{atm}}}.
\label{eq:patmo}
\end{eqnarray}
The energy distribution and flavor ratio of the atmospheric flux are specified by the Honda model~\cite{Honda:2006qj,Sanuki:2006yd} with the Gaisser cosmic ray knee correction~\cite{Gaisser:2013bla}. The probability of passing the atmospheric self-veto $P_{\rm{veto}}$ was calculated using~\cite{Gaisser:2014bja}. 

Finally, the energy dependence of the atmospheric muon likelihood uses the empirical parametrization given by Vincent et al.~\cite{Vincent:2016nut} and the angular distribution from Aartsen et al.~\cite{Aartsen:2015zva}:
\begin{equation}
P_{\mu,i}(t_i,E_i,\vec x_i) = \frac{\int dE_{\mu} R(E,E_{\rm{dep}}) P_{\mu}(E_\mu)P_{\mu}(\vec x)\delta_{t,{\rm{tr}}}}{\int dE_{\rm{dep}} \int d^2\vec{x} \int dE_{\mu} P_{\mu}(E_\mu)P_{\mu}(\vec x)\delta_{t,{\rm{tr}}}},
\end{equation}
where the Kroenecker delta $\delta_{t,{\rm{tr}}}$ ensures only tracks can be counted as atmospheric muons. 

To account for the finite possibility that track events have in fact been misclassified as showers, we perform the following substitution for each observed shower event:
\begin{equation}
P_{i,t = {\rm{sh}}} \rightarrow 0.9 P_{i,t={\rm{sh}}} + 0.1 P_{i,t={\rm{tr}}}.
\end{equation}

\section{Appendix B: Changes in incoming astrophysical power law\label{sec:gamma_var}}
In this appendix we show the results of Fig. 3 
, fixing the spectral index of the extragalactic isotropic flux to three different values: $E^{-2}$, corresponding to the prediction from Fermi acceleration, $E^{-2.6}$, the best fit from \cite{Aartsen:2015zva}, and $E^{-2.9}$ in line with latest fits such as~\cite{Vincent:2016nut,ClaudioTalk}.
\begin{figure}
\begin{tabular}{c c}
\includegraphics[width=0.5\textwidth]{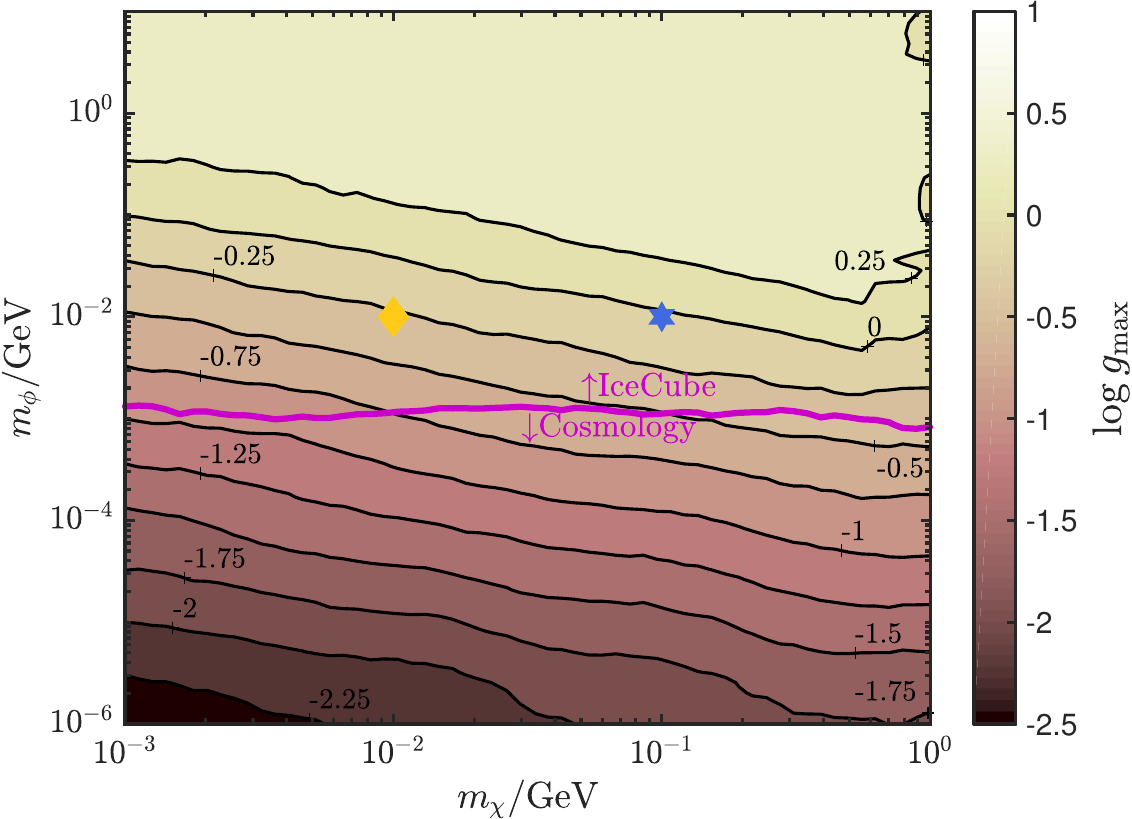}& \includegraphics[width=0.5\textwidth]{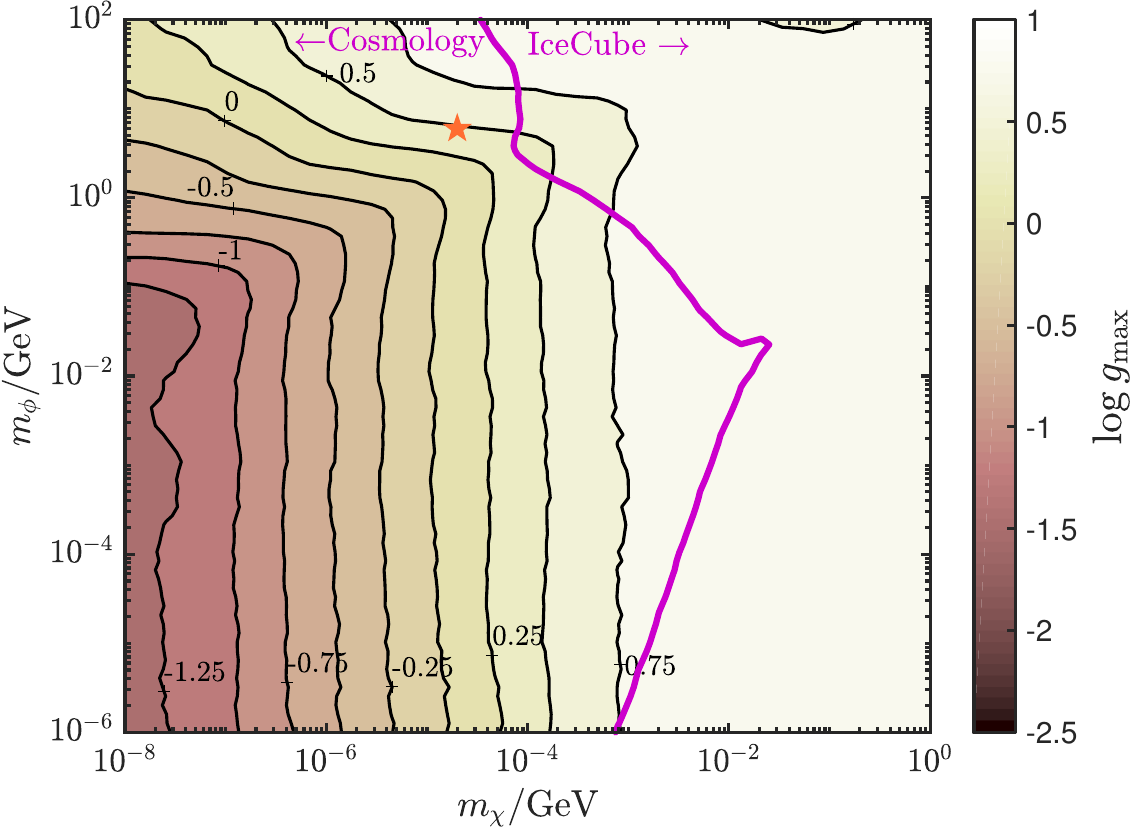} \\
\includegraphics[width=0.5\textwidth]{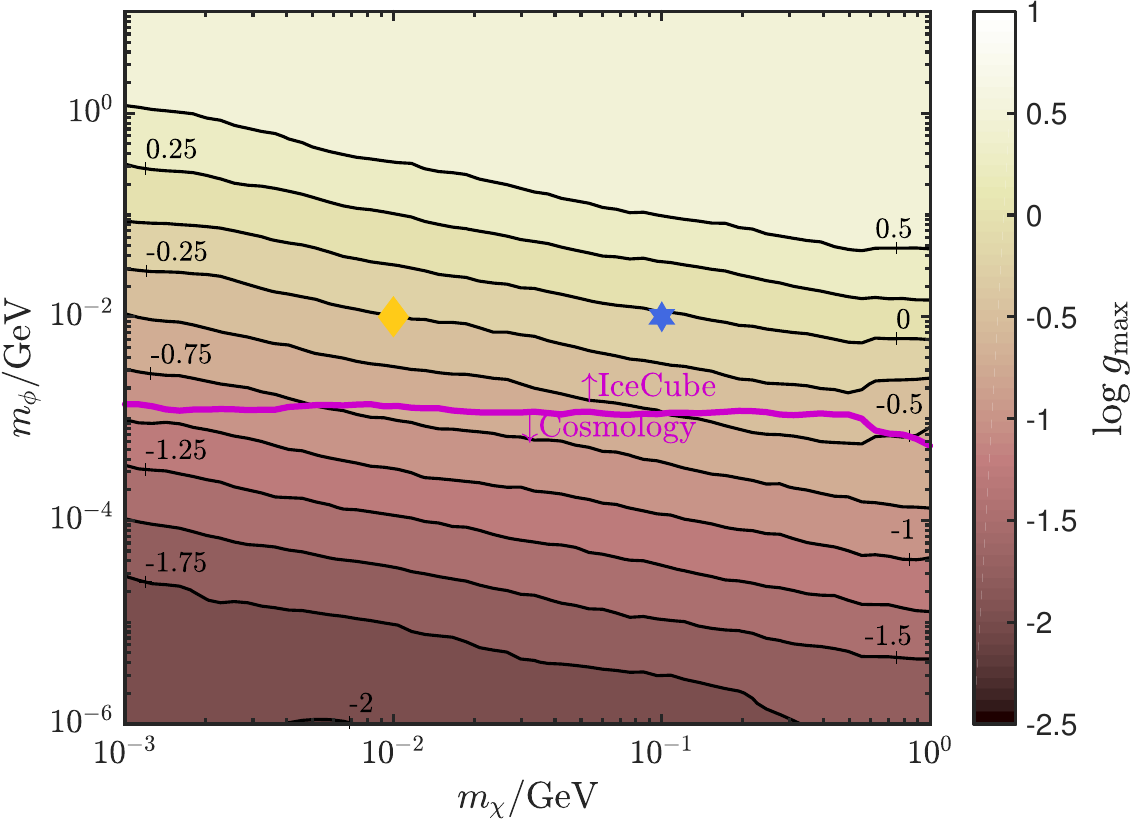}& \includegraphics[width=0.5\textwidth]{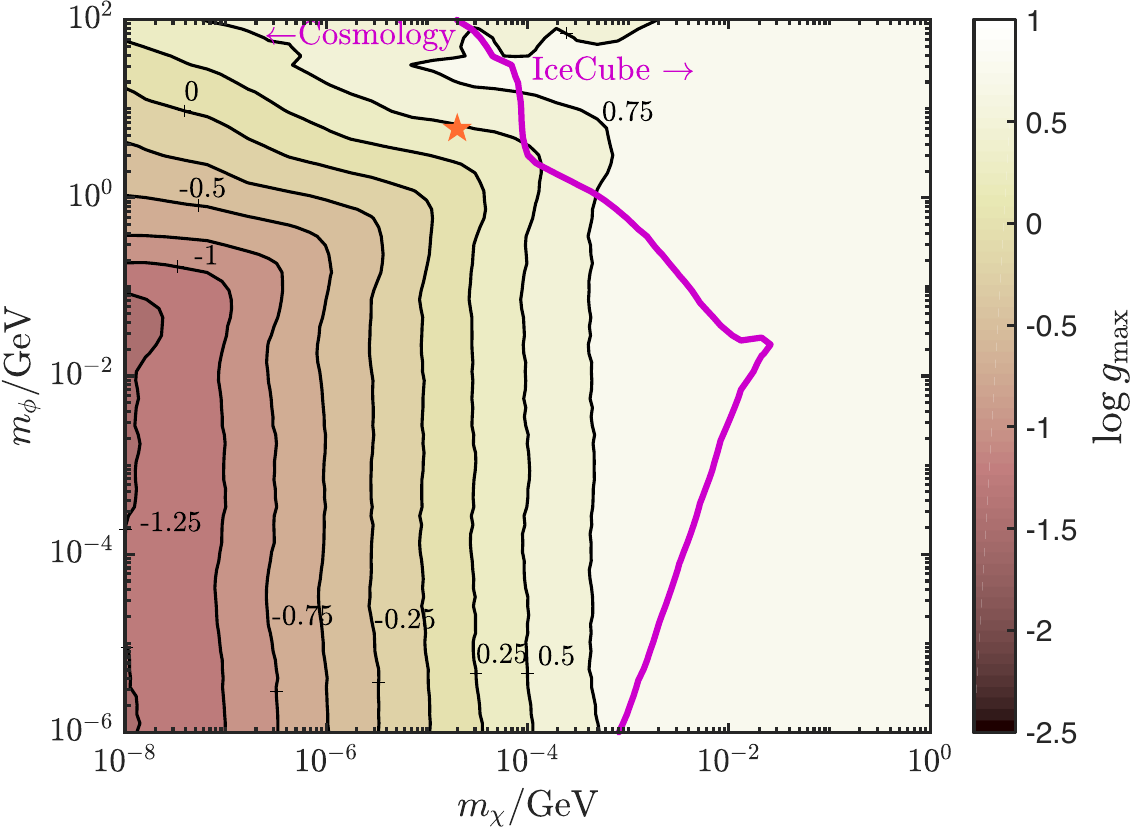} \\
\includegraphics[width=0.5\textwidth]{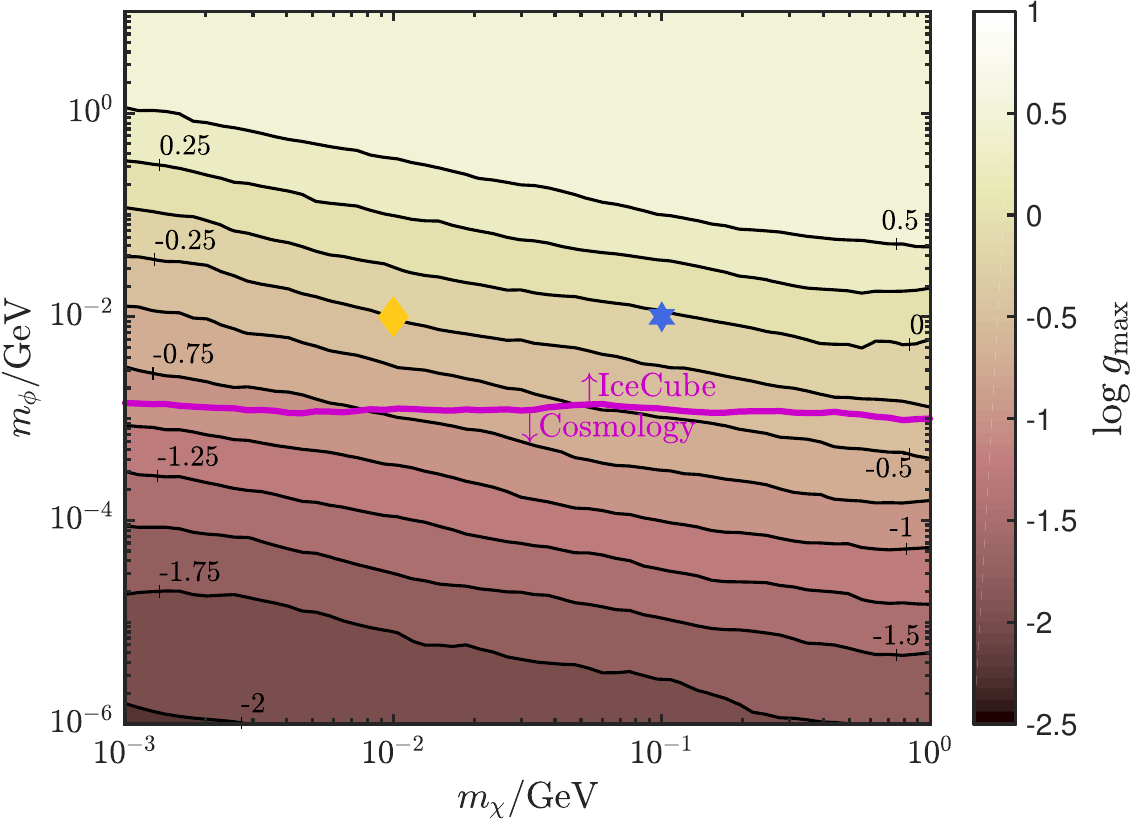}& \includegraphics[width=0.5\textwidth]{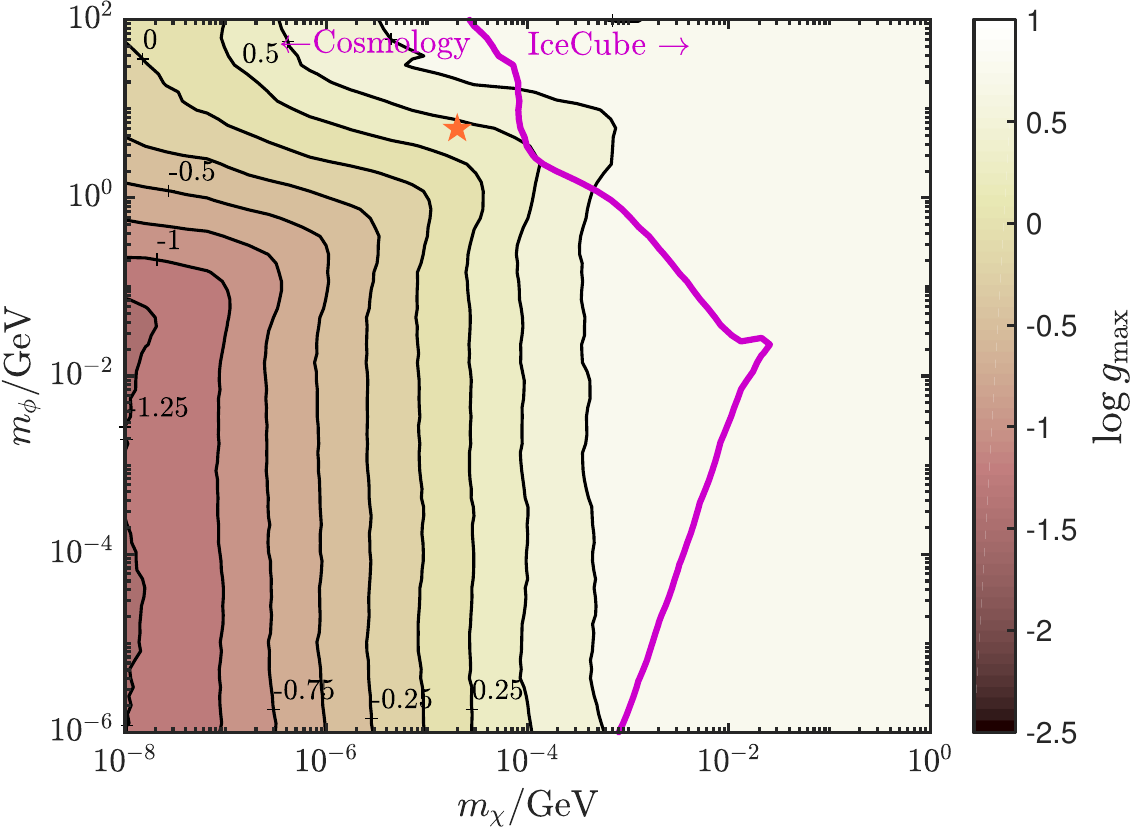} 
\end{tabular}
\caption{Same as Fig. 3
, but for different fixed values of the spectral index $\gamma$ of the isotropic extragalactic flux $\phi_0 \propto E^{-\gamma}$. Top: $\gamma = 2$; middle: $\gamma = 2.6$; bottom: $\gamma = 2.9$. Left: fermionic DM with a vector mediator; right: scalar DM with a fermionic mediator.}
\end{figure}

\section{Appendix C: Cross sections}
Below, we present the dark matter-neutrino interaction cross sections for four simplified models, including those explored here. We thank I. Safa and A.O.~del Campo for independently verifying these. 


We can write the frame-independent expression for the differential cross section:
\begin{equation}
\frac{d \sigma}{d t} = \frac{1}{64\pi s}\frac{1}{|\bold{p}_{1,CM}|^2}|\mathcal{M}|^2.
\end{equation}
In the lab frame, 
\begin{equation}
\frac{d \sigma}{d \cos\theta} = \frac{1}{2 E_\nu} \frac{1}{2 m_\chi} \frac{1}{8\pi}\frac{E_\nu'^2}{E_\nu m_\chi} |\mathcal{M}|^2,
\end{equation}
where the outgoing neutrino energy $E_\nu'$ is related to the incoming energy $E_\nu$ and scattering angle $\theta$ via:
\begin{equation}
E_\nu' = \left(\frac{1}{E_\nu} + \frac{1}{m_\chi}(1-x)\right)^{-1},
\end{equation}
where we have defined $x \equiv \cos \theta$, the scattering angle. In this frame, the Mandelstam variables are:
\begin{eqnarray}
s &=& 2E_\nu m_\chi +m_\chi^2, \nonumber \\
t &=& -2E_\nu E_\nu'(1-x), \nonumber \\
u &=& -2m_\chi E_\nu' + m_\chi^2.
\end{eqnarray}

We can now evaluate the differential and the total cross sections in the lab frame explicitly: \\

\textbf{-- Scalar DM, scalar mediator}: 

\begin{equation}
 \frac{d \sigma}{d \cos\theta} = \frac{g^2 g'^2(1- x) E_{\nu }^2 m_{\chi }}{16 \pi  \left((1-x ) E_{\nu }+m_{\chi }\right) \left((1-x ) E_{\nu } m_{\phi }^2+m_{\chi } \left(m_{\phi }^2-2 (x -1) E_{\nu }^2\right)\right)^2},
\end{equation} 
and
\begin{equation}
 \sigma = -g^2 g'^2\frac{4E_\nu^2m_\chi + (2 E_\nu m_\phi^2 + 4 E_\nu^2 m_\chi + m_\phi^2 m_\chi)\log\left( \frac{m_\phi^2(2 E_\nu + m_\chi)}{2 E_\nu m_\phi^2 + 4 E_\nu^2 m_\chi + m_\phi^2 m_\chi}\right)}{64\pi E_\nu^2m_\chi^2(2E_\nu m_\phi^2 + 4 E_\nu^2 m_\chi + m_\phi^2 m_\chi)}.
\end{equation}

\textbf{-- Dirac DM fermion, scalar mediator}:

\begin{equation}
 \frac{d \sigma}{d \cos\theta} = \frac{g^2 (x-1) \left(g'\right)^2 E_{\nu }^2 m_{\chi }^2 \left(2 (x-1) E_{\nu } m_{\chi }+(x-1) E_{\nu }^2-2 m_{\chi }^2\right)}{8 \pi  \left(m_{\chi }-(x-1) E_{\nu }\right)^2 \left((x-1) E_{\nu } m_{\phi }^2-m_{\chi } \left(m_{\phi }^2-2 (x-1) E_{\nu }^2\right)\right)^2},
\end{equation}
and
\begin{eqnarray}
 \sigma &=&\frac{g^2 \left(g'\right)^2 }{32 \pi  E_{\nu }^2 m_{\chi }^2}\bigg[E_{\nu } m_{\chi }-\frac{E_{\nu } m_{\chi }^2}{2 E_{\nu }+m_{\chi }}-\frac{E_{\nu } m_{\chi }^2 m_{\phi }^2 \left(m_{\phi }^2-4 m_{\chi }^2\right)}{\left(2 E_{\nu } m_{\chi }+m_{\phi }^2\right) \left(4 E_{\nu }^2 m_{\chi }+2 E_{\nu } m_{\phi }^2+m_{\chi } m_{\phi }^2\right)}+\frac{E_{\nu } m_{\chi } \left(m_{\phi }^2-4 m_{\chi }^2\right)}{2 E_{\nu } m_{\chi }+m_{\phi }^2} \nonumber \\
&&+\left(m_{\phi }^2-2 m_{\chi }^2\right) \log  \left(\frac{m_\phi^2(2 E_{\nu }+m_{\chi })}{4 E_{\nu }^2 m_{\chi }+2 E_{\nu } m_{\phi }^2+m_{\chi } m_{\phi }^2}\right)\bigg].
\end{eqnarray}

 \textbf{-- Dirac fermion DM, vector mediator}:

\begin{equation}
 \frac{d \sigma}{d \cos\theta} =\frac{g^2 g'^2 E_\nu^2 m_\chi^2 ((1-x^2)E_\nu m_\chi + (1-x)^2E_\nu^2 + (1+x)m_\chi^2)}{4 \pi  \left((1-x) E_{\nu }+m_{\chi }\right) \left((1-x) E_{\nu } m_{\phi }^2+m_{\chi } \left(m_{\phi }^2-2 (x-1) E_{\nu }^2\right)\right)^2},
\end{equation}
and
\begin{eqnarray}
 \sigma = \frac{g^2 g'^2}{16\pi E_\nu^2 m_\chi^2}&&\left[ (m_\phi^2+m_\chi^2 +2E_\nu m_\chi) \log\left(\frac{m_\phi^2(2E_\nu+m_\chi)}{m_\chi(4E_\nu^2+m_\phi^2)+2E_\nu m_\phi^2}\right)  \right. \nonumber \\  &&+ \left. 4E_\nu^2 \left(1+\frac{m_\chi^2}{m_\phi^2} - \frac{2E_\nu(4E_\nu^2m_\chi + E_\nu (m_\chi^2 + 2 m_\phi^2)+  m_\chi m_\phi^2}{(2E_\nu + m_\chi)(m_\chi(4E_\nu^2 + m_\phi^2 ) + 2E_\nu m_\phi^2)} \right) \right].
\end{eqnarray}

\textbf{-- Scalar DM, fermion mediator}:

\begin{equation}
 \frac{d\sigma}{d\cos\theta} = \frac{g^4 (1+x)E_\nu^4 m_\chi^5((1-x)E_\nu + 2m_\chi)^2}{4\pi (m_\phi^2 - m_\chi(2E_\nu + m_\chi))^2((1-x)E_\nu+m_\chi)^3(E_\nu((x-1)m_\phi^2 - (x+1)m_\chi^2) + m_\chi^3 - m_\chi m_\phi^2)^2},
\end{equation}
and
\begin{eqnarray}
 \sigma =  \frac{g^4}{64\pi} && \left[ \frac{8E_\nu^2 m_\chi}{(2E_\nu+m_\chi)(m_\phi^2-m_\chi(2E_\nu+m_\chi))^2} + \frac{4}{-2E_\nu m_\chi+m_\chi^2-m_\phi^2} +\frac{8}{m_\chi (2E_\nu + m_\chi )-m_\phi^2}     \right. \nonumber \\
&+&\left.  \left(\frac{4E_\nu m_\chi- 2(m_\chi^2+3m_\phi^2)}{E_\nu m_\chi(m_\chi(2E_\nu + m_\chi) - m_\phi^2)} + \frac{3}{E_\nu^2}\right)\log \left(\frac{4E_\nu^2 m_\chi}{m_\phi^2(2E_\nu+m_\chi)-m_\chi^3} +1 \right) \right].
\end{eqnarray}

\end{document}